\documentclass{acmart}

\AtBeginDocument{%
  }

\usepackage{lipsum} 
\usepackage{newfloat}
\usepackage{caption}
\usepackage{multirow}
\usepackage{makecell}
\usepackage{xcolor}
\DeclareFloatingEnvironment[fileext=frm,placement={!ht},name=Frame]{myfloat}

\usepackage{lipsum}
\usepackage[framemethod=TikZ]{mdframed}
\usepackage{hyperref}

\newenvironment{frameenv}[1][]
{\def\mycaption{#1}\begin{myfloat}[tb]
        \begin{mdframed}[roundcorner=10pt,backgroundcolor=white]
        }
        {\end{mdframed}\expandafter\caption{\expandafter\mycaption}\end{myfloat}
}

\usepackage{amsmath}
\begin{document}

\title[Auto-ADMET: Effective and Interpretable AutoML Method for Chemical ADMET Property Prediction]{Auto-ADMET: An Effective and Interpretable AutoML Method for Chemical ADMET Property Prediction}

\author{Alex G. C. de S\'{a}}
\affiliation{%
  \institution{Baker Heart and Diabetes Institute}
  \city{Melbourne}
  \state{Victoria}
  \country{Australia}
  \postcode{3004}
}
\affiliation{%
  \institution{\vspace{0.10cm}School of Chemistry \& Molecular Biosciences, The University of Queensland}
  \city{Brisbane City}
  \state{Queensland}
  \country{Australia}
  \postcode{4067}
}
\affiliation{%
  \institution{Baker Department of Cardiometabolic Health, The University of Melbourne}  
  \city{Parkville}
  \state{Victoria}
  \country{Australia}
  \postcode{3010}
}
\email{Alex.deSa@baker.edu.au}


\author{David B. Ascher}
\affiliation{%
  \institution{Baker Heart and Diabetes Institute}
  \city{Melbourne}
  \state{Victoria}
  \country{Australia}
  \postcode{3004}
}
\affiliation{%
  \institution{School of Chemistry \& Molecular Biosciences, The University of Queensland}
  \city{Brisbane City}
  \state{Queensland}
  \country{Australia}
  \postcode{4067}
}
\affiliation{%
  \institution{Baker Department of Cardiometabolic Health, The University of Melbourne}  
  \city{Parkville}
  \state{Victoria}
  \country{Australia}
  \postcode{3010}
}
\email{d.ascher@uq.edu.au}

\renewcommand{\shortauthors}{de S\'{a} et al.}

\begin{abstract}
Machine learning (ML) has been playing important roles in drug discovery in the past years by providing (pre-)screening tools for prioritising chemical compounds to pass through wet lab experiments. One of the main ML tasks in drug discovery is to build quantitative structure-activity relationship (QSAR) models, associating the molecular structure of chemical compounds with an activity or property.  These properties -- including absorption, distribution, metabolism,  excretion and toxicity (ADMET) -- are essential to model compound behaviour, activity and interactions in the organism. Although several methods exist, the majority of them do not provide an appropriate model's personalisation, yielding to bias and lack of generalisation to new data since the chemical space usually shifts from application to application. This fact leads to low predictive performance when completely new data is being tested by the model. The area of Automated Machine Learning (AutoML) emerged aiming to solve this issue, outputting tailored ML algorithms to the data at hand. Although an important task, AutoML has not been practically used to assist cheminformatics and computational chemistry researchers often, with just a few works related to the field. To address these challenges, this work introduces Auto-ADMET, an interpretable evolutionary-based AutoML method for chemical ADMET property prediction. Auto-ADMET employs a Grammar-based Genetic Programming (GGP) method with a Bayesian Network Model to achieve comparable or better predictive performance against three alternative methods --   standard GGP method, pkCSM and XGBOOST model -- on 12 benchmark chemical ADMET property prediction datasets. The use of a Bayesian Network model on Auto-ADMET's evolutionary process assisted in both shaping the search procedure and interpreting the causes of its AutoML performance. 

\end{abstract}



\maketitle

\section{Introduction}

Artificial Intelligence (AI) and Machine Learning (ML) fields have been empowering drug discovery with predictive methods and tools not only to accelerate its internal pipelines -- e.g., computationally (pre-)screening molecules with adequate properties, such as Absorption, Distribution, Metabolism, Excretion and Toxicity (ADMET) properties~\cite{Mak2023, Serghini2023, Jang2001,Ruiz2008, Hasselgren2024, Yu2010, Pirmohamed2023, deSa2022, Pires2015, Myung2024} -- but also to propose new chemical molecules (e.g., with generative AI, GenAI) -- and, possibly aiming to design new drugs \cite{Arnold2023, Gangwal2024}. Although predictive or generative models might be currently and extensively used to yield new molecules, finding those with correct properties is still challenging and time-consuming. Therefore, while developing customised chemical property prediction models will be a proper way to avoid later obstacles during \emph{in vitro} and \emph{in vivo} testing, it will also be important as these models may be well-integrated into molecular optimisation frameworks because of that \cite{Yu2010, Hasselgren2024, Jang2001, Ruiz2008, Pirmohamed2023}. 

In this work, we primarily focus on ADMET properties of chemical compounds because of their relationships with drug design and optimisation \cite{deSa2022, Pires2015, Xiong2021, Myung2024}. ADMET properties, related to the pharmacokinetics and toxicity of compounds, provide an understanding of how the chemical compounds move into, through, and out of the body. In addition, these properties assess how the molecules affect the human body by assessing how toxic/safe they are to cells, organs and the genome, for example.

Apart from the extensive work on ADMET methods linking the molecule (sub)structure with ADMET properties, most works provide methods, web servers or tools with non-customised and static predictive models. Given the large number of pharmaceutical companies and research institutes working by building compound libraries, we surely have the molecular data drift happening, because of the change of the chemical space. Therefore, it is impractical to assume that these models will maintain their predictive performance in the long run, mainly because the new compounds may be dissimilar to those used to train ADMET property prediction models. With new data, the current solution is to propose new versions of the tools or web servers with updated ADMET models, which is not scalable enough to deal with the demand for novel ADMET property prediction models.  

An efficient alternative to deal with this issue is using Automated Machine Learning (AutoML) methods~\cite{Hutter2019} to recommend customised predictive pipelines to the molecular data at hand. Nevertheless, the majority of previously introduced AutoML methods do not take into account the nature of (bio)chemical data, hence not being able to capture all the steps necessary to succeed in chemical ADMET property prediction tasks, such as molecular representation, molecular data splitting, feature extraction, and data imbalance  \cite{Correia2024}. These steps have been considered in novel AutoML research in the field of AutoML for drug discovery and computational chemistry.

Accordingly, this work proposes Auto-ADMET, an interpretable evolutionary-based AutoML method for chemical ADMET property prediction. Auto-ADMET relies on a Model- and Grammar-based Genetic Programming (GGP) method. While we used a GGP method to evolve valid pipelines in the context of chemical ADMET property prediction tasks, we used a Bayesian Network Classifier (BNC) model in Auto-ADMET to guide and interpret this evolutionary algorithm's decisions.

Twelve (12) chemical ADMET property prediction benchmark datasets were used to validate Auto-ADMET and compare it with alternative methods (i.e., standard GGP method, pkCSM method and XGBOST model). The achieved results of Auto-ADMET on these 12 molecular datasets highlight its predictive power since they show Auto-ADMET's superior performance in eight (8) out of 12 datasets against alternative methods. Moreover, the use of a BNC causal model to assist the GGP algorithm was also relevant to comprehending the AutoML choices during the method's evolution, indicating which algorithms and hyperparameter choices are actually causing the AutoML performance. This understanding of the causes of predictive performance for ADMET prediction might direct efforts to build better methods in future work (e.g., on designing more meaningful AutoML search spaces on the fly).

\section{Related Work}

We divide this section into two subsections. First, we describe evolutionary algorithms that are used to solve computational chemistry problems, which is an area that encompasses chemical ADMET property prediction. Next, we analyse a few works on AutoML related to computational chemistry, focusing on recent works that build and recommend customised predictive pipelines based on (bio)chemical data.

\subsection{Evolutionary Algorithms for Computation Chemistry}

There have been several efforts to solve computational chemistry or cheminformatics problems using evolutionary computation (EC) problems. The survey of Yu et al.~\cite{Yu2024} mainly explores the use of EC for drug discovery, including the development of EC methods for docking, lead compound generation (such as ligands) and exploring the quantitative structure-activity relationships (QSAR) of compounds. 

Nevertheless, this section focuses on using EC for molecular generation and optimisation, and machine learning tasks, which are more related to Auto-ADMET's goals. The works of Soto et al. \cite{Soto2008a} and \cite{Soto2008b} utilise single-objective and multiple-objective genetic algorithms for selecting the best set of descriptors (or features) in ADMET property prediction tasks, respectively. They tested different machine learning models for these studies, including decision trees, k nearest neighbours and polynomic non-linear function regression models to estimate the quality of a given feature set. 

Liu et al. \cite{Liu2018}, in turn, proposed and developed ECoFFeS, which is an evolutionary-based feature selection software designed for drug discovery. ECOFFeS encompasses both single-objective and multi-objective bioinspired or evolutionary algorithms. Whereas ECoFFeS' single-objective algorithms include ant colony optimization (ACO)~\cite{Dorigo1996},  differential evolution (DE)~\cite{Storn1997},  genetic algorithm (GA) \cite{Holland1992} and particle swarm optimization (PSO) \cite{Kennedy1995}, its multi-objective counterpart supports only two well-known Multi-Objective Evolutionary Algorithms (MOEAs) -- i.e., MOEA/D \cite{Zhang2007} and NSGA-II~\cite{Deb2002}.

EC may also be used to optimise chemical compounds, where evolutionary operators can be applied to molecules to derive new ones \cite{Fromer2023}. For example, in Fromer and Coley \cite{Fromer2023}, it is argued that during the evolutionary process aiming to optimise new molecules, a mutation operator might be used to add or remove atoms, bonds or molecular fragments. On the other hand, the crossover operator may be used to exchange molecular fragments among molecules \cite{Fromer2023}. 

In terms of evolutionary computation, molecular optimisation and generation, and large language models, we have the work of da Silva et al. \cite{daSilva2024}, which modelled \emph{de novo} drug design as a many-objective optimization problem (MaOP) since in drug discovery we do have problems with several conflicting objectives (e.g., potency versus safety versus proper pharmacokinetic properties). da Silva et al.'s work involved in developing genAI approaches combined with multi- and many-objective evolutionary algorithms (MOEAs and MaOEAs) for drug development, specifically for combining a generative deep learning model’s latent space with MOEA/MaOEA (NSGA-II/NSGA-III) for designing new and diverse molecules.

\subsection{AutoML for Cheminformatics}

Several ML methods have been proposed for dealing with cheminformatics tasks, including but not limited to pharmacokinetics, human and environmental toxicity, pharmacodynamics and pharmacogenetics \cite{Pires2015, Daina2017, Dong2018, Xiong2021, Cheng2012, Yang2019, deSa2022, Wei2022, Myung2024, Gu2024, Fu2024}. As aforementioned, the main issue in using the ML models derived from these methods is that they are static and non-customisable, leading to possible biases and the lack of predictive generalisation in cases where the input chemical molecules differ from those used to train the ML models. 

Recently, a few works have made efforts to automate and, consequently, personalise cheminformatics or computational chemistry pipelines through search and optimisation, such as the work of de S\'a\ and Ascher \cite{deSa2024}, AutoQSAR \cite{Dixon2016},  ZairaChem \cite{Turon2023}, Uni-QSAR \cite{Gao2023}, QSARtuna \cite{Mervin2024}, and Deepmol \cite{Correia2024}.

de S\'a\ and Ascher \cite{deSa2024} introduced the first evolutionary-based AutoML algorithm to build and recommend tailored predictive pipelines for small molecule pharmacokinetic data. These pipelines included feature definition, scaling and selection, and machine learning algorithms and hyperparameter optimisation. All steps followed by this AutoML algorithm are within a context-free grammar, which is followed to generate individuals, perform genetic operations and guide the evolutionary algorithm to produce only valid solutions. 

AutoQSAR \cite{Dixon2016}, on the other hand, utilises an accuracy score to rank ML pipelines that are aiming to solve a QSAR problem. Nevertheless, AutoQSAR relies on an exhaustive search, not being able to scale in larger datasets. Following a distinct approach, ZairaChem \cite{Turon2023} follows open-source ideas to deliver a robust AutoML package for drug screening, employing five optimisation methods for this purpose independently and targetting different objectives (e.g., predictive performance, interpretability and robustness).

Uni-QSAR \cite{Gao2023} and QSARtuna \cite{Mervin2024} are both automated QSAR frameworks for molecule property prediction. In the case of Uni-QSAR, a stacking method is employed to combine the solutions of several ML models and predict molecule properties as a result. Differently, QSARtuna takes advantage of Bayesian optimisation for the same task.

Finally, Correia et al. \cite{Correia2024} and Li et al. \cite{Li2025} proposed Deepmol and Model Training Engine (MTE), respectively. Both Deepmol and MTE are AutoML frameworks for computational chemistry considering both traditional machine learning and deep learning models. These frameworks are defined by Bayesian optimisation algorithms that search for and optimise pipelines in the context of drug discovery problems. Deepmol's and MTE's search spaces incorporate a list of options, such as standardisation, feature extraction, feature scaling and selection, machine learning modelling and imbalance learning.

\section{Auto-ADMET: AutoML for ADMET Property Prediction}\label{automl}

\begin{figure*}[!htbp]
  \centering
   \includegraphics[scale=0.405]{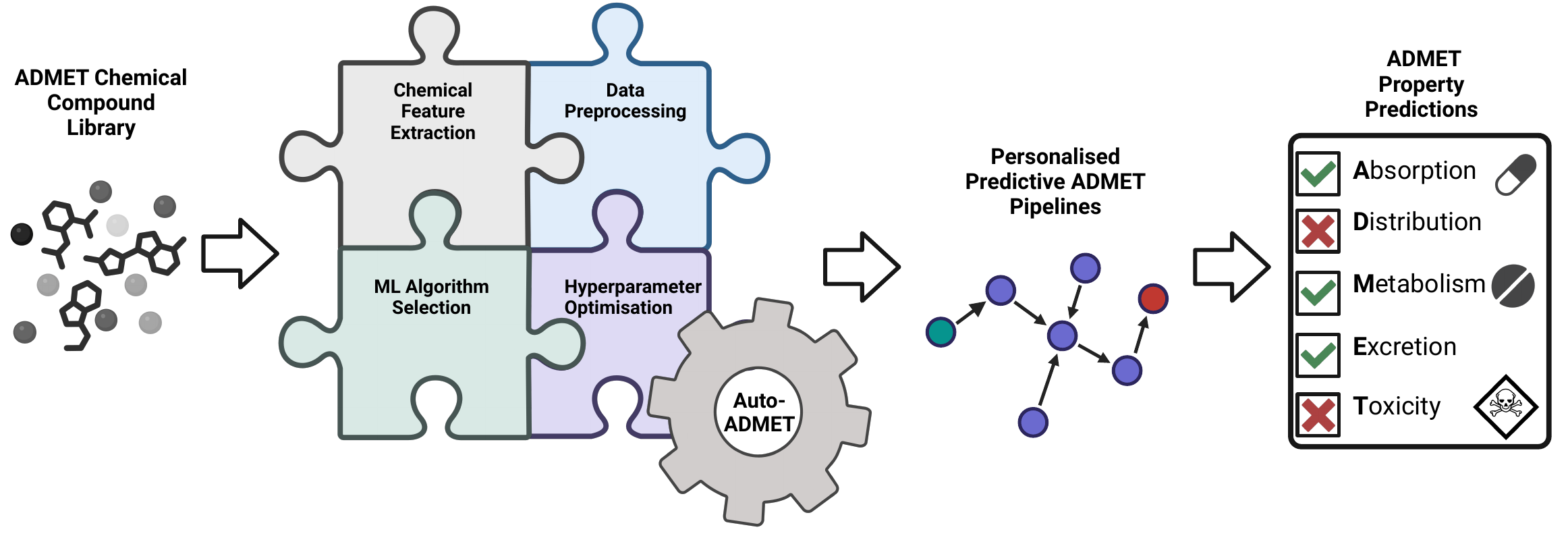}
      \caption{Auto-ADMET's workflow to create personalised machine learning (ML) pipelines targeting ADMET chemical compound property prediction.  }
   \label{fig-main_idea}
 \end{figure*}

Figure \ref{fig-main_idea} illustrates the general workflow followed by Auto-ADMET. It starts by receiving the chemical molecules library targeting ADMET properties as input. Since Auto-ADMET only deals with classification tasks (at the moment), each molecule in the input set contains its associated ADMET labels. Next, Auto-ADMET finds the most suitable combination of ML building blocks for a given ADMET property prediction task, including chemical feature extraction, data preprocessing, ML algorithm selection and hyperparameter optimisation. All these building blocks are automatically selected by Auto-ADMET, which outputs the best-performing and personalised predictive pipeline to the input chemical data. After being fit, this pipeline is able to perform predictions to new molecules\footnote{The official source code for Auto-ADMET is not completely ready, but it will be released soon at \href{https://github.com/alexgcsa/auto-admet}{https://github.com/alexgcsa/auto-admet}.}. 

Next, we present details on the main components of Auto-ADMET. First, we describe Auto-ADMET's search space, highlighting its main ML building blocks, hyperparameters and algorithm options. Second, we introduce the Bayesian network- and evolutionary-based search method, which employs the search space to find and optimise ML pipelines in the context of ADMET prediction tasks.
 
\subsection{Search Space}

A context-free grammar defines the \emph{search space} employed by Auto-ADMET, comprehending five (5) chemical extraction methods (and all of their combinations), five (5) scaling techniques (and the possibility of not using any scaling on the dataset), six (6) feature selection approaches (and the possibility of not using any feature selection on the dataset) and 10 machine learning algorithms.

The excerpt of the grammar defining Auto-ADMET's search space can be found in Frame \ref{frame1}. The grammar is formally presented as a four-tuple <N, T, P, S>, where: N is a set of non-terminals; T is a set of terminals; P is a set of production rules; and S (a member of N) is the start symbol. The production rules in the grammar derive the language by combining the grammar symbols. In addition,  the symbol "|" represents a choice, and the non-terminals surrounded by the symbols "[" and "]" are optional, i.e., they can appear or not in the production rules.

The start rule—<Start> in the grammar shown in Frame \ref{frame1}--defines the four main components of the ADMET prediction pipeline: (i) molecular representation (captured by the non-terminal <feature\_definition>), (ii) feature\ scaling, (iii) feature\ selection, and (iv) machine learning modelling (represented by the non-terminal <ML\_algorithms>).

For molecular representation -- which relates to the chemical feature extraction step, 31 different combinations of chemical compound representation techniques are available. These fall into five main categories: molecular descriptors, advanced molecular descriptors, graph-based signatures, fragments, and toxicophores \cite{Pires2015, deSa2022}. This \emph{search space} component essentially determines the features used to characterise compounds based on their biochemical structure.

Feature scaling is handled using standard approaches from the scikit-learn library \cite{Pedregosa2011, Raschka2019}, including Normalizer, Min-Max Scaler, Max Abs Scaler, Robust Scaler, and Standard Scaler. This step modifies the numerical representation of chemical compounds, ensuring that feature values are appropriately scaled. However, the grammar also allows the option of bypassing feature scaling, as specified in the <Start> rule.

Feature selection is another key component involving the selection of relevant features using methods from scikit-learn \cite{Raschka2019}. The grammar includes Variance Threshold, Select Percentile, Recursive Feature Elimination (RFE) and selection methods based on False Discovery Rate (FDR), False Positive Rate (FPR), and Family-Wise Error Rate (FWE). It is worth noting that RFE requires a predictive model on top of the feature selection method. As with scaling, the grammar provides the flexibility to proceed without applying any feature selection method, as defined in the <Start> rule.

The grammar is designed exclusively for classification pipelines \cite{Raschka2019}. At present, ML modelling consists of 10 algorithms implemented in scikit-learn and other independent software \cite{Pedregosa2011, Chen2016}: Decision Tree, Extremely Randomized Tree (Extra Tree), Random Forest, Adaptive Boosting (AdaBoost), Gradient Boosting, Neural Networks (using Multi-Layered Perceptron), Support Vector Machines (SVM), Nu-Support Vector Machines (NuSVM) and Extreme Gradient Boosting (XGBoost).

For all main categories, hyperparameter optimisation is also applied to each level, ensuring the search of their values to the input data. Considering all available choices and their respective hyperparameters, the AutoML grammar comprises 59 non-redundant production rules, with 58 non-terminals and 389 terminals.

  \begin{frameenv}[The excerpt of the proposed AutoML grammar.\label{frame1}]
       <Start> ::= <feature\_definition> [<feature\_scaling>] [<feature\_selection>] <ML\_algorithms> \vspace{0.1cm}
       
       <feature\_definition> ::= General\_Descriptors | Advanced\_Descriptors |
                       Graph-based\_Signatures | Toxicophores |
                       Fragments | General\_Descriptors Advanced\_Descriptors | General\_Descriptors Graph-based\_Signatures | 
                       ... | General\_Descriptors Advanced\_Descriptors
                       Graph-based\_Signatures Toxicophores 
                       Fragments \vspace{0.3cm}
                       
        <feature\_scaling> ::= <Normalizer> | <MinMaxScaler> | <MaxAbsScaler> | <RobustScaler> | <StandardScaler> \vspace{0.1cm}

        <Normalizer> ::= Normalizer <norm>
        
        <norm> ::= l1 | l2 | max\vspace{0.1cm}
        
        ... \vspace{0.3cm}

        <StandardScaler> ::= StddScaler <with\_mean> <with\_std>

        <with\_mean> ::= True | False
        
        <with\_std> ::= True | False \vspace{0.3cm}

        <feature\_selection> ::= <Variance\_Threshold> | <Select\_Percentile> | <Select\_FPR> | <Select\_FWE> | <Select\_FDR> | <Select\_RFE>\vspace{0.1cm}
        
        <Variance\_Threshold> ::= VarianceThreshold <threshold>

        <threshold> ::= 0.0 | 0.05 | 0.10 | 0.15 | ... | 0.85 | 0.90 | 0.95 | 1.0\vspace{0.1cm}
        
        ... \vspace{0.3cm}
        
        <ML\_algorithms> ::= <AdaBoost> | <Decision\_Tree> | <Extra\_Tree> | <Random\_Forest> | <Extra\_Trees> | <Gradient\_Boosting> | <Neural\_Networks> | <SVM> | <NuSVM> | <XGBoost>\vspace{0.1cm}

        <AdaBoost> ::= AdaBoost <algorithm> <n\_estimators> <learning\_rate>

        <algorithm> ::= SAMME.R | SAMME
        
        <n\_estimators> ::= 5 | 10 | 15 | 20 | ...| 300 | 500 | 550 | ... |  950 | 1000 | 1500 | 2000 | 2500 | 3000
        
        <learning\_rate> ::= 0.01 | 0.02 | 0.03 | ... | 2.0\vspace{0.1cm}
        
        ... \vspace{0.3cm}

        <XGBoost> ::= XGBoost <n\_estimators> <max\_depth> <max\_leaves> <learning\_rate>
        
        <max\_depth> ::= 1 | 2 | 3 | 4 | 5 | 6 | 7 | 8 | 9 | 10 | None
        
        <max\_leaves> = 1 | 2 | 3 | 4 | 5 | 6 | 7 | 8 | 9 | 10
    \end{frameenv}

\subsection{Search Method}

Inspired by the Bayesian Optimisation Algorithm (BOA) \cite{Pelikan2005}, Auto-ADMET makes use of both the Grammar-based Genetic Programming (GGP) method and a Bayesian Network Classifier (BNC) to help the GGP search to explore and exploit more promising areas of the search space.

\begin{figure}[!htbp]
  \centering
   \includegraphics[scale=0.35]{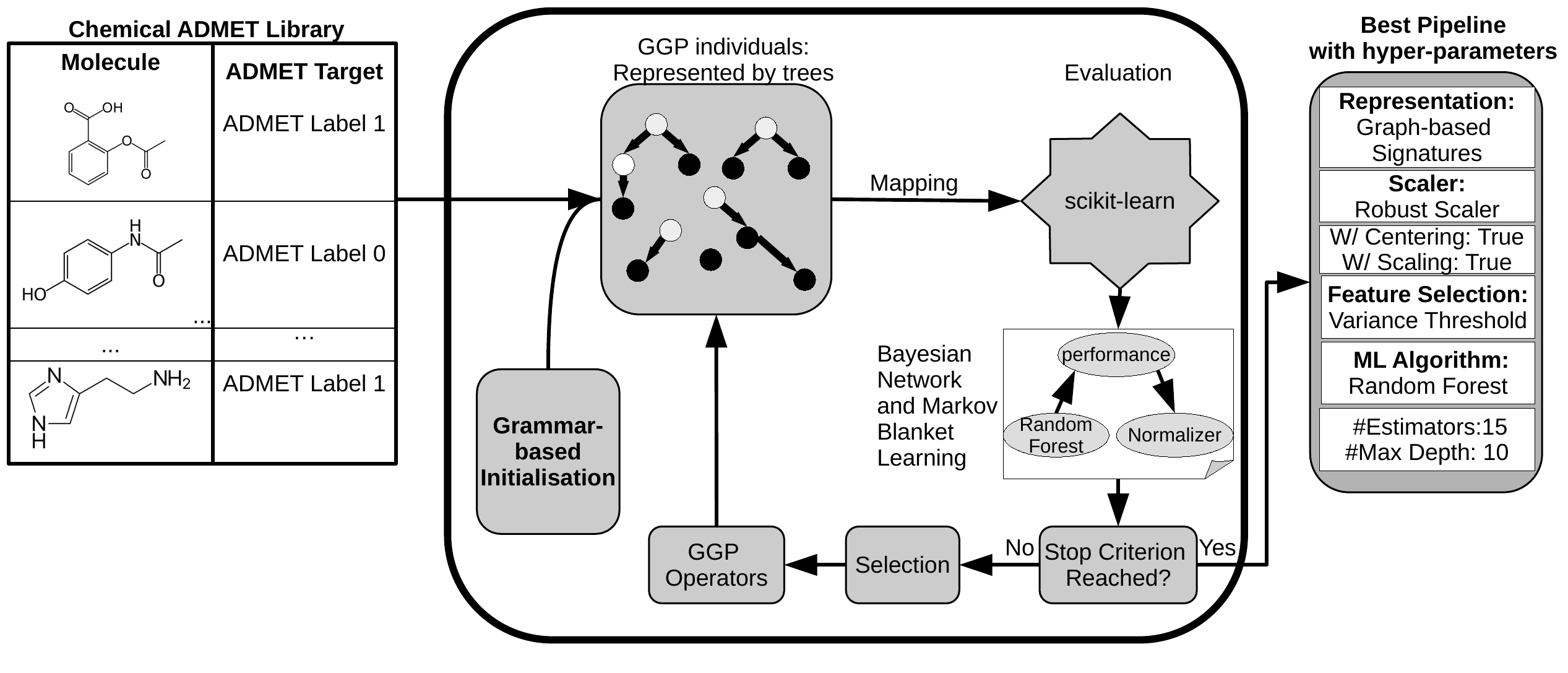}
   \caption{The grammar-based genetic programming (GGP) method with a Bayesian Network Classifier to search for ML pipelines in the context of ADMET property prediction. Figure adapted from \cite{deSa2017, deSa2024}.}
   \label{fig-general}
 \end{figure}

Figure \ref{fig-general} illustrates the workflow followed by Auto-ADMET. First, a chemical dataset containing compounds is set as input to Auto-ADMET. This dataset could correspond to any particular chemical compound property, including but not limited to absorption, distribution, metabolism, excretion and toxicity (ADMET) properties.

Next, the GGP method gets its population of individuals (in our case, individuals are machine learning pipelines) initialised at random but following the grammar rules expressed in the previous section. All individuals are converted into scikit-learn pipelines to be evaluated in it. Provided the results of the evaluation (see Section \ref{fitness} for more details), we train a Bayesian Network Classifier (BNC) \cite{Friedman1997} taking into account the main building blocks in feature extraction, feature scaling, feature selection and ML algorithm modelling. The target of the BNC is the actual performance of classification pipelines, where a threshold of 80\% and 60\% of the current best predictive performance is used to set good and bad pipelines, respectively. 

From the BNC, we take its Markov Blanket\footnote{In a Bayesian network, the Markov Blanket of a node is the set of the nodes corresponding to the node's parents, its children and its co-parents} and use it to sample in areas of the \emph{search space} that are actually causing the performance. To build the BNC, we used a Hill Climbing with a Bayesian Information Criterion as the scoring metric. We used the aGrUM/pyAgrum's implementation for this step \cite{Ducamp2020} to train and build the Bayesian Network.

Based on this, a proportion of the new population will have only algorithms sampled based on the BNC's Markov Blanket. The next step involves checking whether the stopping criterion has been met. If it has not, the pipelines undergo selection based on the fittest individuals based on tournament selection and the application of GGP operators, specifically Whigham’s crossover and mutation, and crossover followed by mutation~\cite{Mckay2010}. If both operators are used, mutation modifies the recombined pipelines resulting from the crossover operation. Importantly, both crossover and mutation adhere to the grammar constraints, ensuring that only valid individuals are generated. Additionally, elitism is employed, preserving the top $n$ pipelines from the previous generation to maintain high-performing solutions.

This evolutionary process, illustrated in Figure \ref{fig-general}, continues iteratively until the stopping criterion is satisfied. Once met, the best-performing pipeline from the final evaluated population is returned, along with the optimal hyperparameters identified through the GGP method with the BNC model.

\subsection{Fitness Function}\label{fitness}

A similar evaluation process to de S\'a and Ascher \cite{deSa2024}'s work is used here. However, instead of employing traditional cross-validation, Auto-ADMET relies on nested 3-fold cross-validation with three (3) trials to ensure the correct fitness of the pipeline.

To assess the quality of each pipeline, the \emph{fitness function} is defined as the average of Matthew’s correlation coefficient (MCC)~\cite{Chicco2020} over the employed cross-validation procedure. MCC is a widely used performance metric in classification tasks, particularly valuable in cases of data imbalance due to its robustness in evaluating model performance, which is the case for chemical datasets. The MCC formula is given in Equation~\ref{MCC}:

\begin{equation} \label{MCC} MCC = \frac{((TP \times TN) - (FP \times FN))}{\sqrt{(TP + FP) \times (TP + FN) \times (TN + FP) \times (TN + FN)}} \end{equation}

In this equation, TP (true positives) refers to molecules labelled as one (1) and correctly predicted as one (1), TN (true negatives) represents molecules labelled as zero (0) and correctly predicted as zer (0), FP (false positives) denotes molecules labelled as zero (0) but incorrectly predicted as one (1), and FN (false negatives) corresponds to molecules labelled as one (1) but incorrectly predicted as 0 (zero).

MCC accounts for all four types of classification outcomes (TP, TN, FP, and FN), providing a balanced evaluation even in imbalanced learning scenarios. As a correlation coefficient, MCC ranges from -1.0 to +1.0, where +1.0 indicates a perfect positive correlation, -1.0 represents a complete inverse correlation, and an MCC of 0.0 suggests that the classifier's predictions are no better than random guessing.

\section{Experiments}

This section outlines the key aspects of the AutoML experiments for ADMET chemical data, including a description of the datasets (Section \ref{datasets}), the configuration of the grammar-based genetic programming (GGP) method with the Bayesian Network Classifier (Section \ref{configuration}), and the benchmarking with alternative approaches (Section \ref{comparisons}).

\subsection{Chemical ADMET Datasets}\label{datasets}

Twelve (12) chemical ADMET datasets were used to evaluate the performance of Auto-ADMET. These datasets represent binary classification tasks related to absorption, metabolism, and excretion based on experimental \emph{in vivo} or \emph{in vitro} tests on small chemical molecules. The number of molecules (\# Molecules) in these datasets varies significantly, ranging from 404 to 18,558. This variation in dataset size presents a challenge for Auto-ADMET, requiring it to effectively adapt and optimise the search process for different pipeline configurations. Table \ref{tab:datasets} encompasses the employed datasets and their characteristics. 

\begin{table}[!htbp]

    \centering
    \caption{Description of 12 binary classification ADMET datasets.}
    \begin{tabular}{|l|p{2.5cm}|l|l|l|}
        \hline
        \textbf{ID} & \textbf{Dataset} & \textbf{Abbreviation} & \textbf{Category} & \textbf{\# Molecules} \\
        \hline
        1 & Caco-2 permeability & Caco-2 & Absorption & 663 \\ \hline
        2 & P-glycoprotein I Inhibitor & PGP I Inhibitor & Absorption & 1223 \\ \hline
        3 & P-glycoprotein II Inhibitor & PGP II Inhibitor & Absorption & 1023 \\ \hline
        4 & P-glycoprotein I Substrate & PGP I Substrate & Absorption & 1272 \\ \hline
        5 & Skin Permeability & Skin Perm. & Absorption & 404 \\ \hline
        6 & Cytochrome P450 CYP2C9 Inhibitor & CYP2C0 Inhibitor & Metabolism & 14,706 \\ \hline
        7 & Cytochrome P450 CYP2C19 Inhibitor & CYP2C19 Inhibitor & Metabolism & 14,572 \\ \hline
        8 & Cytochrome P450 CYP2D6 Inhibitor & CYP2D6 Inhibitor & Metabolism & 14,738 \\ \hline
        9 & Cytochrome P450 CYP2D6 Substrate & CYP2D6 Substrate & Metabolism & 666 \\ \hline
        10 & Cytochrome P450 CYP3A4 Inhibitor & CYP3A4 Inhibitor & Metabolism & 18,558 \\ \hline
        11 & Cytochrome P450 CYP3A4 Substrate & CYP3A4 Substrate & Metabolism & 669 \\ \hline
        12 & Renal Organic Cation Transporter 2 Substrate & OCT2 Substrate & Excretion & 904 \\
        \hline
    \end{tabular}
    \label{tab:datasets}
\end{table}

It is important to emphasise that although we selected these 12 datasets to validate Auto-ADMET, any other chemical predictive task would be suitable to be applied as an Auto-ADMET input. We plan to assess the performance of Auto-ADMET in a broad range of datasets in future work.

The complete datasets were divided into two sets: training data and blind test data, using a stratified approach. Specifically, 90\% of each dataset was allocated for searching the optimal pipeline for the respective ADMET dataset, while the remaining 10\% was reserved to evaluate the accuracy of the final selected pipeline.

\subsection{Parameter Configuration}\label{configuration}

The GGP (with BNC) parameters were configured in the following way. 100 individuals representing ML pipelines are evolved for one hour. Each individual has at most 5 minutes to run. Otherwise, its run is interrupted, and its score is multiplied by 0.7. Crossover mutation operators are employed with a probability rate of 0.0 and 0.15, respectively. Crossover followed by mutation is applied at a rate of 0.05. Over the generations, the best current individual is kept for the next generations (i.e., the elitism size is equal to 1). From the individuals sampled from the BNC's Markov Blanket, we select a maximum of 10\% of the population size (i.e. 10) of new individuals (ML pipelines) to compose the new population. Finally, to avoid overfitting happening on the final model generated by the best pipeline, we add a random pipeline into the population if we have cases where 70\% of the ML pipelines in the population are the same, indicating convergence. Table \ref{tab:configurations} highlights these Auto-ADMET parameters.

\begin{table}[htbp]

    \centering
    \caption{The GGP parameters for evolving a population of machine learning pipelines for PK prediction.}
    \begin{tabular}{|l|l|}
        \hline
        \textbf{Parameter} & \textbf{Value} \\
        \hline 
        Population Size & 100 \\ \hline
        Stopping Criterion & 1 hour \\ \hline
        Crossover Probability & 0.80 \\ \hline
        Mutation Probability & 0.15 \\     \hline   
        Crossover and Mutation Probability & 0.05 \\ \hline     
        Elitism Size & 1 \\ \hline
        BNC sampled population &  10 (10\%) \\ \hline
        Individual's time budget & 5 minutes \\ \hline
        Population similarity rate to add random individuals & 70\% \\        
        \hline
    \end{tabular}
    \label{tab:configurations}
\end{table}

\subsection{Benchmarking} \label{comparisons}

We evaluated the best pipelines discovered by Auto-ADMET (Section \ref{automl}) by comparing them against three alternative approaches. First, after running Auto-ADMET 20 times--to ensure reliable statistical analysis--and collecting the resulting pipelines, we assessed their performance using a 3-fold nested cross-validation procedure. The best AutoML-discovered pipeline was then compared to pkCSM \cite{Pires2015}, a widely recognised method for predicting the chemical ADMET properties.

Second, we compared the best-selected pipelines and the best AutoML-optimised pipeline against XGBoost \cite{Chen2016} with its default parameters. XGBoost was chosen due to its widespread use in machine learning and its frequent application in predictive modelling.

Moreover, we assessed the best pipelines found by Auto-ADMET by comparing them against those discovered by the standard GGP method proposed by de S\'a and Ascher \cite{deSa2024}. Its results are also included in the following section.

Finally, to statistically compare all four (4) methods, we applied Iman-Davenport’s modification of Friedman’s test~\cite{Demvsar2006}. If the test yielded significant results, we conducted a Nemenyi \emph{post hoc} test to perform pairwise comparisons, assessing whether the predictive performance of the AutoML method and its best-found pipeline significantly differed from the alternative approaches.

\section{Results}

This section presents the results considering the experiments detailed in the previous section. We first provide a comparison in terms of predictive performance in Section \ref{comparison} among all the methods (among best pipelines found on cross-validation) -- the work of de S\'a and Ascher~\cite{deSa2024},  pkCSM \cite{Pires2015}, XGBoost \cite{Chen2016} and Auto-ADMET. Next, we analyse a case of study on the dataset \emph{Caco-2} to understand how the evolutionary process is taking advantage of the Bayesian Network Classifier to evolve machine learning pipelines better.

\subsection{Benchmarking against Alternative Methods} \label{comparison}

When performing the analysis of the benchmarking of Auto-ADMET against alternative methods (the AutoML method proposed by de S\'a and Ascher (2024), pkCSM and XGBOOST) in Table \ref{tab:comparison}, we noticed that Auto-ADMET achieves the highest average MCC (0.618) across all datasets, outperforming de Sá and Ascher (0.530), pkCSM (0.456), and XGBoost~(0.497).
In terms of average ranking, Auto-ADMET is also ranked as the best method (1.667), followed by de Sá and Ascher (2024) (1.917), pkCSM (3.333) and XGBoost (3.083).

When looking at dataset by dataset in Table \ref{tab:comparison}, we observed the good behaviour of Auto-ADMET in providing machine learning pipelines that handle well ADMET predictive tasks. In summary, Auto-ADMET outperforms all other methods in eight (8) out of 12 datasets (Caco-2, PGP II Substrate, Skin Permeability, CYP2C9 Inhibitor, CYP2C19 Inhibitor, CYP2D6 Substrate, CYP3A4 Inhibitor and OCT2 Substrate).
It has a particularly strong lead in CYP3A4 Inhibitor (0.958), where the second-best method (pkCSM) achieves only 0.623.

\begin{table}[htbp]
    \centering
    \caption{Comparison of the best proposed AutoML method found by Auto-ADMET, across the 20 runs against the best pipeline found by the work of de S\'a and Ascher (2024), pkCSM and XGBOOST (XGB) in terms of MCC metric on the blind test set.}
    \begin{tabular}{|c|l|c|c|c|c|}
        \hline
        \textbf{ID} & \textbf{\makecell{Dataset}} & \textbf{\makecell{Auto-ADMET}} & \textbf{\makecell{de S\'a and Ascher (2024) \cite{deSa2024}}} & \textbf{pkCSM} & \textbf{XGB} \\
        \hline
        1 & \makecell{Caco-2} & 0.641 & 0.610 & 0.609 & 0.579 \\  \hline
        2 & \makecell{PGP I \\ Inhibitor} & 0.799 & 0.837 & 0.776 & 0.820 \\ \hline
        3 & \makecell{PGP II \\ Inhibitor} & 0.703  & 0.783 & 0.716 & 0.696 \\ \hline
        4 & \makecell{PGP II \\ Substrate} & 0.367 & 0.289 & 0.214 & 0.232 \\ \hline
        5 & \makecell{Skin \\ Perm.} & 0.588 & 0.394 & 0.108 & 0.368 \\ \hline
        6 & \makecell{CYP2C9 \\ Inhibitor} & 0.668  & 0.615 & 0.601 & 0.553 \\ \hline
        7 & \makecell{CYP2C19 \\ Inhibitor} & 0.764 & 0.647 & 0.583 & 0.590 \\ \hline
        8 & \makecell{CYP2D6 \\ Inhibitor} & 0.540 & 0.556 & 0.408 & 0.488 \\ \hline
        9 & \makecell{CYP2D6 \\ Substrate} & 0.581 & 0.334 & 0.197 & 0.267 \\ \hline
        10 & \makecell{CYP3A4 \\ Inhibitor} & 0.958 & 0.590 & 0.623 & 0.534 \\ \hline
        11 & \makecell{CYP3A4 \\ Substrate} & 0.145 & 0.274 & 0.289 & 0.440 \\ \hline
        12 & \makecell{OCT2 \\ Substrate} & 0.661 & 0.427 & 0.353 & 0.402 \\
        \hline
        \multicolumn{2}{|c|}{\textbf{Average}} & 0.618 & 0.530 & 0.456 & 0.497 \\ \hline
        \multicolumn{2}{|c|}{\textbf{Ranking}} & 1.667 & 1.917 & 3.333 & 3.083\\
        \hline
    \end{tabular}
    \label{tab:comparison}
\end{table}

When performing a statistical test to compare the methods, the Friedman Test (with Iman-Davenport Correction) reports a statistically significant difference (p-value = 0.00047), indicating that at least one method performs significantly differently from the others. Therefore, we proceed with a \emph{post hoc} analysis with the Nemenyi Test for pairwise comparisons in accordance with the critical difference. The critical difference (1.4072) suggests that Auto-ADMET’s superiority over pkCSM and XGBoost is statistically meaningful. We also identified that the pipelines found by the work of de S\'a and Ascher perform better than pkCSM with statistical difference.

Although we are constantly improving Auto-ADMET to reach better AutoML predictive performance against alternative approaches, its main proposal is not only to deliver good results but also to explain them. With a Bayesian Network Classifier (BNC) guiding the decisions of Auto-ADMET's evolutionary process, we claim Auto-ADMET may actually model AutoML performance more wisely. We analyse this aspect next in Section \ref{analysis}

\subsection{Auto-ADMET's interpretation through Bayesian Network Analysis} \label{analysis}

\begin{figure*}[!htbp]
  \centering
   \includegraphics[scale=0.75]{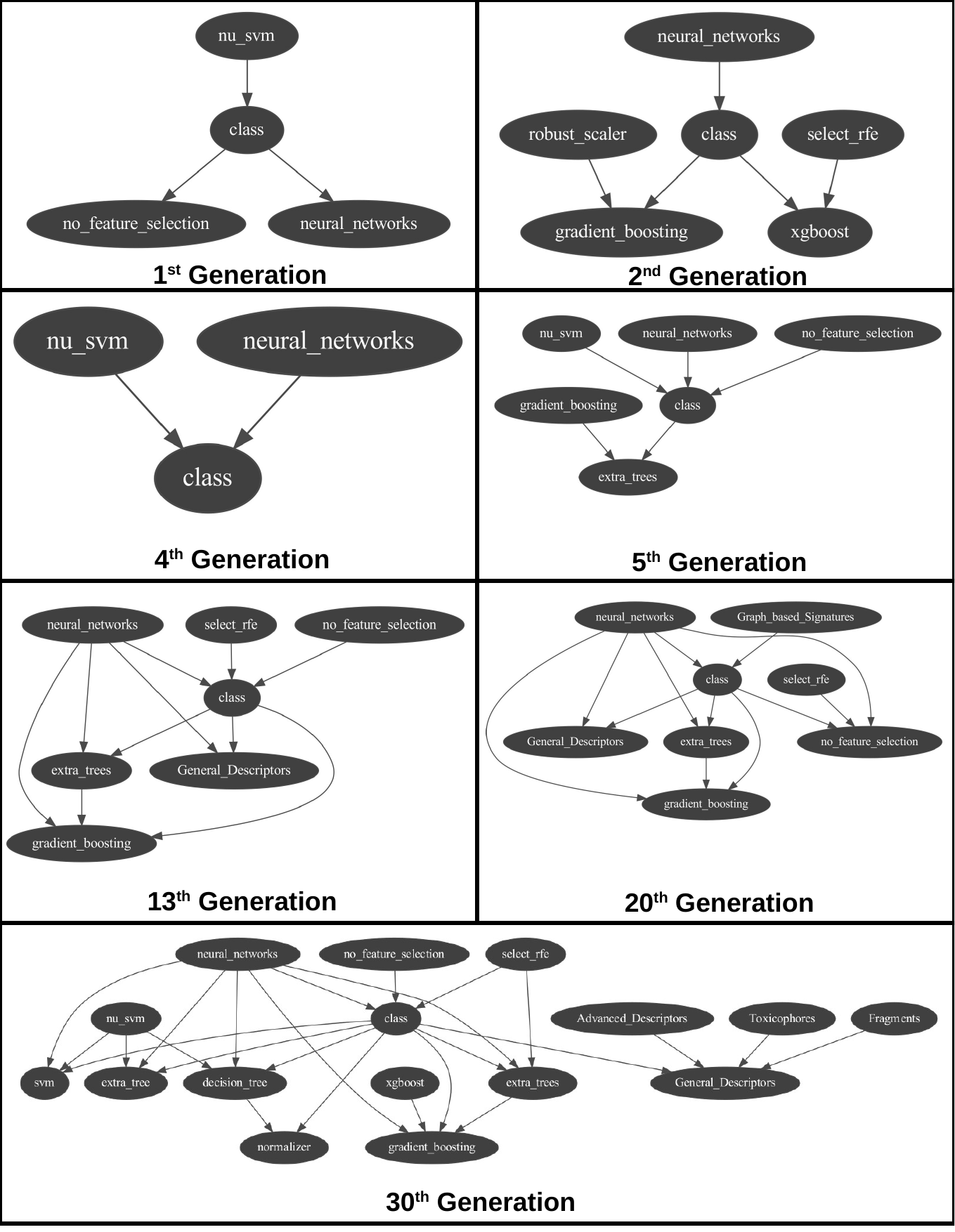}
      \caption{The built Bayesian Network Classifiers across Auto-ADMET's generations. }
   \label{fig-BNCs}
 \end{figure*}

Figure \ref{fig-BNCs} depicts the Markov Blankets of the Bayesian Network Classifiers (BNCs) throughout the evolutionary process followed by Auto-ADMET. Apart from using these BNCs to sample from regions of the \emph{search space} that are more prominent to cause the performance, they can also be used to understand the decisions the evolutionary algorithm (i.e., grammar-based genetic programming) is (partially) making over time.

We can observe from Figure \ref{fig-BNCs} that the BNC also evolve from simpler to more complex, as data regarding machine learning pipeline evaluations targetting ADMET properties is increasing over time. For example, in the first generation, performing feature selection is not an aspect that causes performance, as opposed to the second generation. However, there is an alternation of pipeline components over time. The 30th generation shows the power of machine learning algorithms to cause performance, as most of the nodes of the Markov blanket of the BNC are actually algorithm components, even if using or not using feature selection (Recursive Feature Elimination and No Feature Selection are present in the 30th generation).

 \section{Conclusions and Future Work}

This paper introduces Auto-ADMET, a novel, robust and interpretable AutoML method for predicting chemical absorption, distribution, metabolism, excretion and toxicity (ADMET) properties. While Auto-ADMET pays attention to recommending good predictive pipelines composed of molecular representation, scaling, feature selection and ML modelling, one of its main goals is also to have its decisions easily explained by an evolutionary algorithm guided by a Bayesian Network Classifier.

Preliminary results on 12 ADMET datasets demonstrate Auto-ADMET's capabilities in selecting and configuring cheminformatics pipelines for ADMET predictive tasks, although not limited to them. The performed analysis on Auto-ADMET's results against alternative methods in terms of Matthew's correlation coefficient put Auto-ADMET as the best method in outputting predictive pipelines that will yield good models and results.

Nevertheless, although these results indicate a good step forward in proposing new specific AutoML methods for cheminformatics and ADMET problems, we still plan to improve the search and optimisation methods in Auto-ADMET to translate them to improved predictive performance. In fact, one of our further studies is to design the Bayesian Network Classifier differently. For instance, instead of interactively and locally building the Bayesian Networks across the evolutionary process, we can actually perform a complete study \emph{a priori} and use it to build a better causation model to guide Auto-ADMET's evolution.

Moreover, we expect to compare Auto-ADMET to similar methods in future work, such as ZairaChem \cite{Turon2023}, Uni-QSAR~\cite{Gao2023}, QSARtuna~\cite{Mervin2024}, Deepmol \cite{Correia2024} and Model Training Engine (MTE) \cite{Li2025}. With these comparisons, we would be able to understand where Auto-ADMET is at in terms of AutoML predictive performance. In this evaluation, we will standardise the comparison by using the chemical datasets found in the ADMET group of Therapeutics Data Commons~\cite{Huang2021}.

We trust the evolutionary algorithm assisted with the Bayesian Network to model performance causation will benefit reiterations to model the AutoML \emph{search space}, and therefore, lead to both improved AutoML \emph{search space} and \emph{search method} designs. Alternative ideas on how to model predictive performance and its causation effects are also targeted for future work.

\section*{Funding}
Investigator Grant from the National Health and Medical Research Council of Australia (GNT1174405); Victorian Government’s Operational Infrastructure Support Program (in part).

\bibliographystyle{ACM-Reference-Format}
\bibliography{abbrev.bib}

\end{document}